%% file: main.tex
\documentclass[conference]{IEEEtran}
\usepackage{color}
\usepackage{cite}
\usepackage[normalem]{ulem}
\usepackage{graphicx}
\usepackage{psfrag}
\usepackage{subfigure}
\usepackage{url}
\usepackage{stfloats}
\usepackage{amsmath}
\usepackage{amsthm,mathrsfs,amsfonts,dsfont} 
\usepackage{array}
\usepackage{epsfig}
\usepackage{setspace}
\usepackage{amssymb}
\usepackage{diagbox}
\usepackage{slashbox}

\usepackage{algorithm}
\usepackage{algorithmic}

\usepackage{booktabs}

\usepackage{bm}

\usepackage{comment}

\input{macro}

\begin{document}

\title{Dual Gradient Descent EMF-Aware MU-MIMO Beamforming in RIS-Aided 6G Networks}
\author{\IEEEauthorblockN{Yi Yu, Rita Ibrahim and Dinh-Thuy Phan-Huy}
\IEEEauthorblockA{
 Orange Labs\\
92320, Châtillon, France \\
Email: yu.yi@orange.com, rita.ibrahim@orange.com, dinhthuy.phanhuy@orange.com
}}
\maketitle

\begin{abstract}
Reconfigurable Intelligent Surface (RIS) is one of the key technologies for the upcoming 6th Generation (6G) communications, which can improve the signal strength at the receivers by adding artificial propagation paths. In the context of Downlink (DL) Multi-User Multiple-Input Multiple-Output (MU-MIMO)  communications, designing an appropriate Beamforming (BF) scheme to take full advantage of this reconfigured propagation environment and improve the network capacity is a major challenge.
Due to the spatial dimension provided by MIMO systems,  independent data streams can be transmitted to multiple users simultaneously on the same radio resources.
It is important to note that serving the same subset of users over a period of time may lead to undesired areas where the average Electromagnetic Field Exposure (EMFE) exceeds regulatory limits.  
To address this challenge, in this paper, we propose a Dual Gradient Descent (Dual-GD)-based Electromagnetic Field (EMF)-aware MU-MIMO BF scheme that aims to optimize the overall capacity under EMFE constraints in RIS-aided 6G cellular networks.

\end{abstract}

\begin{IEEEkeywords}
Dual gradient descent, EMF exposure, MU-MIMO, RIS, Reinforcement learning, 6G networks. 
\end{IEEEkeywords}

\section{Introduction} 
\label{sec:motivation}
\IEEEPARstart{6}{G} has enormous commercial potential and is attracting attention from both academia and industry \cite{OrangeSite}. Various innovative technologies are being extensively studied for application in the 6G era. One of these hot topics is the Reconfigurable Intelligent Surface (RIS), which is essentially a large array of low-cost passive components that performs  phase shift of incident waves to reflect them in the desired direction \cite{RIS}. In this way, additional propagation paths can be artificially added to reconfigure the propagation environment and improve the link budgets between transmitters and receivers.

Meanwhile, in the 6G era, operators will continue to leverage the Multi-User Multiple-Input Multiple-Output (MU-MIMO) technology \cite{pinchera2021optimizing} with massive MIMO (M-MIMO) antennas to meet increasing data rate demands. Downlink (DL) MU-MIMO technology enables efficient spatial multiplexing by applying appropriate Beamforming (BF) weights that direct signals to target devices and mitigate or eliminate the influence of interfering data streams.

However, sometimes the radiation patterns generated by DL MU-MIMO BF may produce some undesired areas of strong Electromagnetic Field Exposure (EMFE). The International Commission on Non-Ionizing Radiation Protection (ICNIRP) \cite{icnirp} has specified the average limits of human exposure to Electromagnetic Field (EMF) for a given time period \cite{baracca2018statistical}. These EMFE limits are habitually respected due to some averaging factors met in the network such as scheduling decision, traffic demand, users' spatial distribution etc. However, respecting EMFE limits becomes more challenging \cite{gsma} when the same subset of users is served for long periods of time, such as in fixed wireless access use cases.

Therefore, it is crucial to deploy in the network an efficient EMF-aware MU-MIMO BF that meets the high requirements of 6G networks. Dual Gradient Descent (Dual-GD) is an iterative Reinforcement Learning (RL) algorithm, which can cope with optimization problems under multiple linear inequality constraints. This is suitable for our problem: designing a MU-MIMO BF scheme that maximizes the network capacity under maximum transmit power constraint and EMFE constraints on all the observation points. The key idea of the Dual-GD technique is transforming the original constrained optimization problem into a Lagrange dual function which can be optimized iteratively. This algorithm involves an alternation between maximizing the Lagrangian function with respect to the primal variables and decrementing the Lagrange multipliers by theirs gradients. By repeating this iteration, we can gradually adjust the Lagrangian multiplier corresponding to each constraint according to its impact on the optimization objective, and the solution will converge.

In \cite{awarkeh2021electro,DT1RIS,DT2RIS,xu2019analysis}, the authors have proposed different EMF-aware BF schemes in RIS-aided Single-User MIMO (SU-MIMO) networks. In \cite{ourRIS2022}, we have focused on the RIS-aided MU-MIMO scenario and proposed two EMF-aware BF schemes: (i) "reduced" EMF-aware BF which consists of decreasing the overall transmit power until the EMFE limits are fulfilled and (ii) "enhanced" EMF-aware BF with a per-layer power control mechanism. In this paper, we refine these findings and propose a novel Dual-GD based EMF-aware MU-MIMO BF scheme that enhances furthermore the network capacity while strictly satisfying EMFE constraints.

The rest of the paper is organized as follows. A 6G MU-MIMO RIS-aided network model is defined in section \ref{sec:model}. The "reference" BF scheme (i.e.  without any EMFE constraint) is presented in section \ref{sec:reference}. We describe the details of the Dual-GD EMF-aware BF scheme in the  section \ref{sec:beamforming}. Then, in section \ref{sec:numerical}, the performance of the Dual-GD BF scheme in terms of DL channel capacity and power efficiency is evaluated and compared with other MU-MIMO BF schemes  (i.e. "reference", "reduced" and "enhanced"). Finally, the paper is concluded in section \ref{sec:conclusion}.

\section{System Model}
\label{sec:model} 

\begin{figure}[htbp]
\centering 
\includegraphics[width = 0.8\linewidth]{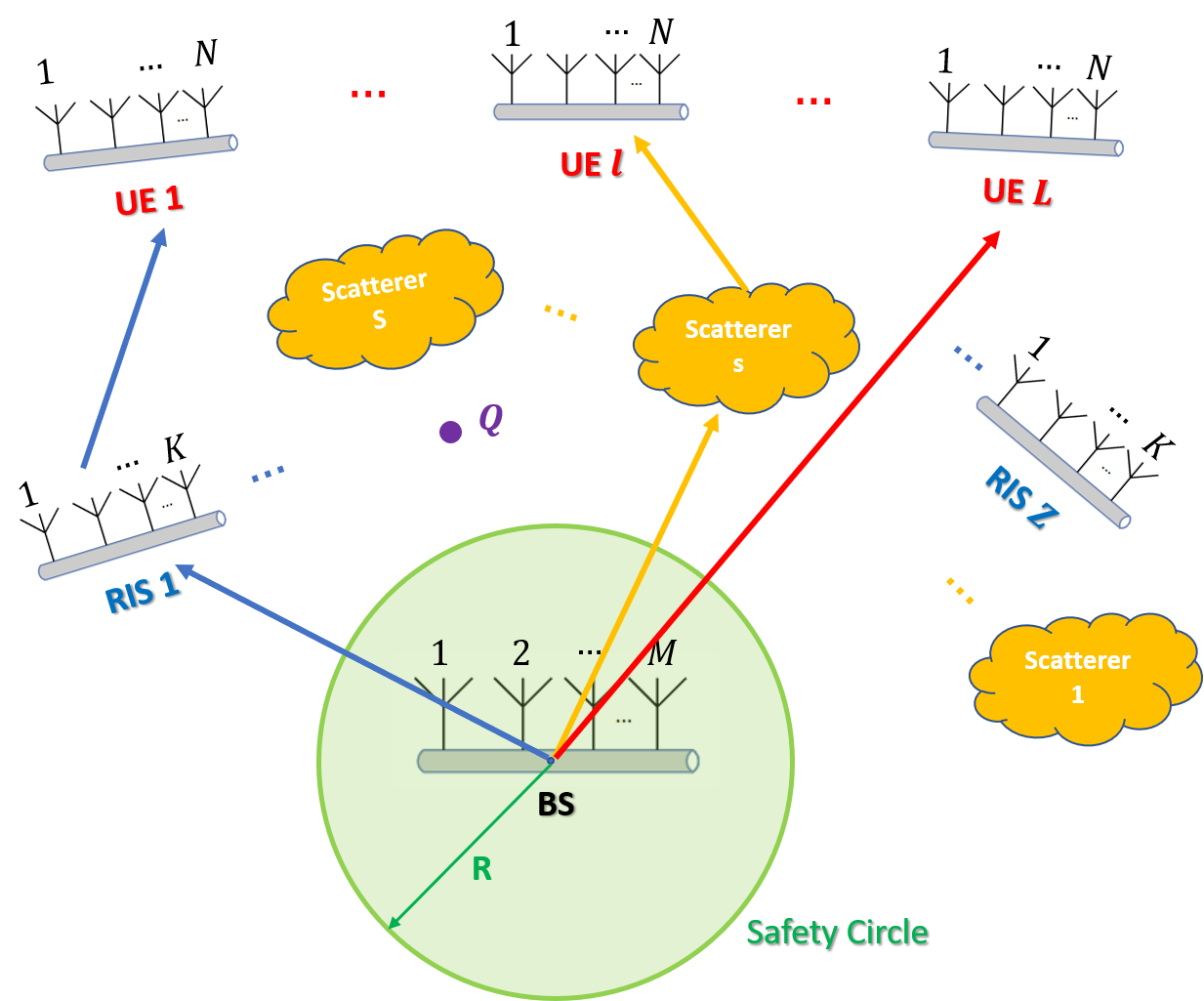}
\caption{A MU-MIMO RIS-aided Network Model}
\label{fig:network}
\end{figure}

In this section, we consider the DL MU-MIMO communications of a RIS-aided cellular network. As shown in Fig. \ref{fig:network}, we assume a single cell scenario with $L$ different User Equipment (UEs). The Base Station (BS) is equipped with a 2D antenna array of $M$ transmitting antenna elements. According to the  3GPP standard \cite{3GPPRelease17}, the BS is modeled by a uniform rectangular panel array, with $N_H$ the number of columns and $N_V$ the  number of antenna elements with the same polarization in each column. We assume that the antenna panel is dual polarized (i.e. $P=2$). So $M = N_H  N_V P$. Both the horizontal $d_H$ and vertical $d_V$ antenna spacing are equal to $0.5\lambda$, where $\lambda$
indicates the wavelength of the carrier frequency. Each UE is equipped with $N$ receiving antenna elements spaced by $0.5\lambda$. The total number of received antennas is thus $N_t=LN$.  

Assume that $S$ scatterers and $Z$ RISs are randomly distributed in the given cell space. Each RIS is equipped with a linear array of $K$ elements spaced by $0.5\lambda$. Both scatterers and RISs are assumed far from the BS and the UEs, therefore for simplicity, we consider the far-field calculation method, i.e. the electromagnetic waves propagate at the speed of light and electric and magnetic fields are mutually perpendicular \cite{baracca2018statistical}. 

We consider an Orthgonal Frequency Division Multiplexing (OFDM) waveform and random Rayleigh fading. 
The network adopts Time Division Duplex (TDD) mode and thus the channel reciprocity is feasible. With MU-MIMO, multiple streams are sent from the BS to distinct active UEs simultaneously. These streams are spatially multiplexed by using appropriate BF schemes. In our work, an adapted channel inversion BF is applied: Zero Forcing (ZF) precoding  adapted to multiple receiving antennas scenario. 

In this paper, the main focus is on the BF at the BS side. The joint optimization of the BS and RIS BF weights is the subject of our future work. Here, we assume that the RISs
are randomly distributed as reflective surfaces to work on transmitting the incident signal to a specific UE. The reflection weights at the RIS side are selected based on the following procedure:

\begin{itemize}
 \item Each UE sends some pilots which allow the RIS to estimate the UE-to-RIS channels;
  \item Based on the UE-to-RIS channel estimation, each RIS computes the BF reflection weight $\mathbf{w}^{z} \in \mathbb{C}^{K \times 1}$;
  \item The weight $\mathbf{w}^{z} $ is multiplied by a reflection amplitude $r^{ris}$, where $0 \leq r^{\text {ris }} \leq 1$ is a constant value depending on the hardware structure of the RIS. Here we set $r^{r i s}=1 / K$; 
  \item Then each RIS applies these reflection weights and freezes;
\end{itemize}
Once the RISs are configured, the UE sends pilots again in such a way that the BS can estimate the DL channel taking into account the RIS configuration and determine the appropriate BF weight to be used for data transmission.

In order to satisfy the EMF exposure compliance, a safety circle of radius $R$ centered at the BS is defined. Outside this safety circle, the received power at any location within the observation range should not exceed a given threshold $\text{EMF}_{\text {th }} \in \mathbb{R}^{+}$. The safety circle, also known as the exclusion zone, is guaranteed to be closed to the public.

In our network model, there are three different kinds of propagation paths:

\begin{enumerate}
\item $m \rightarrow U_{n}^{l}$ denotes direct Line of Sight (LoS) propagation from the $m^{th}$ BS antenna element to the $n^{th}$ antenna element of the $l^{th}$ UE.
  \item $m \rightarrow s \rightarrow U_{n}^{l}$ indicates the path from the $m^{th}$ antenna element of the BS to the $n^{th}$  antenna element of the $l^{th}$ UE through scatterer $s$ .
  \item $m \rightarrow R^z_k \rightarrow U_{n}^{l}$ is the path from $m^{th}$ BS antenna element to the $n^{th}$ antenna element of the $l^{th}$ UE, through $k^{th}$ antenna element of the $z^{th}$ RIS.
  
\end{enumerate}
with $1 \leqslant m \leqslant M$, $1 \leqslant n \leqslant N$, $1 \leqslant l \leqslant L$, $1 \leqslant k \leqslant K$, $1 \leqslant s \leqslant S$ and $1 \leqslant z\leqslant Z$.



According to the 3GPP standardization \cite{3GPPRelease17}, the 3D antenna radiation pattern of each antenna element in the horizontal cut is generated as:
\begin{equation}
A_{\mathrm{dB}}\left(\theta=90^{\circ}, \phi\right)= -\min \left\{12\left(\frac{\phi}{\phi_{3\mathrm{dB}}}\right)^{2}, A_{\max }\right\},
\end{equation}
with $\phi_{3\mathrm{dB}}=65^{\circ}$, $A_{\max }=30 \mathrm{~dB}$ and $\phi \in\left[-180^{\circ}, 180^{\circ}\right]$ is the azimuth angle.

In case of polarized antennas, the polarization is modeled as angle-independent in both azimuth and elevation. In the horizontal polarization, the antenna element field component is given by:
\begin{equation}
F_{\theta,\phi}=\sqrt{A_{\text {Beam }}(\theta, \varphi)} \sin (\zeta),
\end{equation}
with $\zeta = +/- 45^{\circ}$ being the polarization slant angle corresponds to a pair of cross-polarized antenna elements. For the detailed calculation of the 3D radiation pattern $A_{\text {Beam }}(\theta, \varphi)$ of the entire antenna array, please refer to Appendix \ref{app:antenna}.

The propagation channel $\boldsymbol{H}_{l} \in \mathbb{C}^{N \times M}$ between the BS and a given UE $l$ through the considered scatterers and RISs is modeled by:
\begin{equation}
\boldsymbol{H}_{l}[n,m]= G_{m, U_{n}^{l}} + \sum_{s=1}^{S} G_{m, s, U_{n}^{l}} +\sum_{z=1}^{Z} \sum_{k=1}^{K} G_{m, R_{k}^{z}, U_{n}^{l}},
\end{equation}
where $G_{m, U_{n}^{l}}$, $G_{m, s, U_{n}^{l}}$ and $G_{m, R_{k}^{z}, U_{n}^{l}}$ are the channel gains of paths $m \rightarrow U_{n}^{l}$, $m \rightarrow s \rightarrow U_{n}^{l}$ and  $m \rightarrow R^z_k \rightarrow U_{n}^{l}$, respectively. For the calculation of these channel gains, please refer to Appendix \ref{app:derivation}.


Hence, the combined channel matrix $\boldsymbol{H}$ is written as:


$$\boldsymbol{H}=\left[\begin{array}{l}\boldsymbol{H}_{1} \\ \boldsymbol{H}_{2} \\ \cdots \\ \boldsymbol{H}_{L}\end{array}\right] \in \mathbb{C}^{N_t \times M}.$$

Assume that the propagation between the BS and a random nearby position $Q \in \mathbb{R}^{3 \times 1}$ is free space propagation,  $\boldsymbol{H}_{m}^{Q} $ is the $m^{th}$ coefficient of the channel model $\boldsymbol{H}^{Q} \in \mathbb{C}^{1 \times M}$.
Here $\boldsymbol{H}_{m}^{Q}$ is calculated by:
\begin{equation}
\boldsymbol{H}_{m}^{Q}=F^{\prime}_{\theta,\phi} \cdot \frac{\lambda e^{-j \frac{2 \pi}{\lambda}\left\|\overrightarrow{A^{BS}_{m} Q}\right\|}}{4 \pi\left\|\overrightarrow{A^{BS}_{m} Q}\right\|},
\end{equation}
where  $F^{\prime}_{\theta,\phi}$ is the 3GPP radiation power pattern corresponds to the spherical angles $(\theta,\phi)$ of a given path. $F^{\prime}_{\theta,\phi}$ is converted to a linear scale, where $F^{\prime}_{\theta,\phi} =10^{ F_{\theta,\phi}/10}$. $A^{BS}_m\in \mathbb{R}^{3 \times 1}$ is the position of the $m^{th}$ antenna element of the BS.

In our MU-MIMO system, the data vector is denoted by $\boldsymbol{x}=\left[\boldsymbol{x}_{1}^T, \boldsymbol{x}_{2}^T, \cdots, \boldsymbol{x}_{L}^T\right]^T \in \mathbb{C}^{\nu \times 1}$ with $\nu_l$ layers transmitted to the $l^{th}$ UE ($\boldsymbol{x}_{l} \in \mathbb{C}^{\nu_l \times 1}$) and $\nu$ is the total number of spatial layers $\nu=\sum_{l=1}^{L} \nu_{l}$. The components of the data vectors are normalized, i.e. $\mathbb{E}\left[{\|\boldsymbol{x}\|}^2\right]=1$. The data $\boldsymbol{x}$ should be pre-processed via the ZF precoder with the BF matrix denoted as $\boldsymbol{B} = [\mathbf{B}_{1}\cdots\mathbf{B}_{L}]\in \mathbb{C}^{M \times \nu}$. 
Then, the final transmitted signal $\boldsymbol{s} \in \mathbb{C}^{M \times 1}$ is given by:
\begin{equation}
\boldsymbol{s}= \boldsymbol{B}\boldsymbol{x}.
\end{equation} 

The received signal vector is,
\begin{equation}
\boldsymbol{y}= \boldsymbol{H}\boldsymbol{B}\boldsymbol{x}+\boldsymbol{n},
\end{equation}
where $\boldsymbol{n}$ indicates the random receiving noise.

\section{Reference MU-MIMO BF Scheme (ZF precoding with water-filling power allocation)}
\label{sec:reference}

In this section, without considering the EMFE constraints, we adopt a ZF precoding based BF scheme with water-filling power control to help the BS to transmit the signals under the total transmit power constraint $P_{max}$. More specifically, since the component of the data vectors are normalized, i.e. $\mathbb{E}\left[\boldsymbol{x}\boldsymbol{x}^{H}\right]=1$, the power constraint is expressed as follows:
\[
\mathbb{E}\left[ {\|\boldsymbol{Bx}\|}^2\right]=tr\left[ \boldsymbol{B}\boldsymbol{B}^{H}\right] \leqslant P_{max},
\]


The SVD of each full-rank channel matrix $\boldsymbol{H}_{l} \in \mathbb{C}^{N \times M}$ corresponding to UE $l$ is given by: 
\begin{equation}
\boldsymbol{H}_l =\boldsymbol{U}_l\boldsymbol{\Lambda}_l  \boldsymbol{V}_{l}^{H},  
\label{eq:SVD}
\end{equation}
where $\boldsymbol{U}_l\in \mathbb{C}^{N \times N}$ and $\boldsymbol{V}_l\in \mathbb{C}^{M \times N}$ are respectively the unitary orthogonal matrices representing the subset of the left-singular and  right-singular vectors. $\boldsymbol{V_l}^{H}$ represents the conjugate transpose of $\boldsymbol{V_l}$.
$\boldsymbol{\Lambda}_l=diag\left\{\sqrt{\lambda_{l,1}}, \cdots, \sqrt{\lambda_{l,N}} \right\} $ is a ${N \times N}$ diagonal matrix containing the singular vectors of the channel matrix $\boldsymbol{H}_l$.

Taking into account the receiving diversity at the UE level, the ZF BF matrix $\boldsymbol{B}$ is determined based on the pseudo-inverse of the concatenated matrix $\boldsymbol{V} = \left[\mathbf{V}_{1},   \mathbf{V}_{2}, \ldots, \boldsymbol{V}_{L}\right]^H \in \mathbb{C}^{N_t \times M}$, where the pseudo-inverse matrix $\boldsymbol{V}^{+}\in \mathbb{C}^{M \times N_t}$ is given by:
\begin{equation}
\boldsymbol{V}^{+} = \boldsymbol{V}^H \left(\boldsymbol{V} \boldsymbol{V}^{H}\right)^{-1}.
\label{eq:pseudo}
\end{equation}



In our scenario, we pick only some layers that we are interested in, e.g., $\nu_l \leqslant rank(\boldsymbol{H_l})$ layers per receiver. Then the total number of layers is equal to $\nu = \sum_{l =1}^{L}\nu_l$ with $\nu \leqslant N_t$.  With only $\nu$ layers being selected, the matrix $\boldsymbol{V}^{+}$ is trimming to  $\widetilde{\boldsymbol{V}}^{+} \in \mathbb{C}^{M \times \nu}$.
A total of $\nu$ vertical columns corresponding to the different selection layers are retrieved from $\boldsymbol{V}^+$ and reconstituted into this $\widetilde{\boldsymbol{V}}^+$ matrix.

The BF matrix $\boldsymbol{B}\in \mathbb{C}^{M \times \nu}$ is then deduced as:
\begin{equation}
\begin{aligned}
\boldsymbol{B} =\widetilde{\boldsymbol{V}}^{+}  \boldsymbol{\Sigma}, 
\end{aligned}
\end{equation}
with $\boldsymbol{\Sigma} \in \mathbb{C}^{\nu \times \nu}$ being a diagonal power allocation matrix. 
The transmit power coefficient of a selected layer is set to $P_i$, respectively, where $i\in [1,\cdots,\nu]$. Therefore, $\boldsymbol{\Sigma}$ is denoted as,

\begin{equation}
 \begin{array}{l}
 \boldsymbol{\Sigma}=
 diag\{\underbrace{\sqrt{P_{1}}, \cdots, \sqrt{P_{\nu_1}}}_{\nu_{1}}, \cdots,
 \underbrace{\sqrt{P_{\nu-\nu_L}}, \cdots, \sqrt{P_{\nu}}}_{\nu_{L}} \} .
 \end{array}
 \end{equation}






As mentioned previously, the total transmit power is bounded by $tr\left[ \boldsymbol{B}\boldsymbol{B}^{H}\right]\leqslant P_{max}$. So we have, 

$$
tr\left[ \boldsymbol{B}\boldsymbol{B}^{H}\right] =tr\left[\boldsymbol{\Sigma}^2    \left(\widetilde{\boldsymbol{V}}\widetilde{\boldsymbol{V}}^{H}\right)^{-1}  \right] \leqslant P_{max}.
$$



Thanks to ZF precoding, the interference between different users is reduced and the DL capacity of the MU-MIMO system is approximated by:
\begin{equation}
C=\omega \, \sum_{i=1}^{\nu} \log \left(1+\frac{\lambda_i  P_i}{N_{0}}\right),
\text { Mbits\,/\,s},
\label{eq:C}
\end{equation}
where  $\omega$ represents the bandwidth and $N_0$  is  the  power  density  of  the  noise. 

To achieve the maximum data rate, we are going to find the transmit power allocation that satisfies this optimization expression:

\begin{equation}
\mathrm{C}^{*}:=\max _{P_{1}, \ldots, P_{\nu}} \omega \, \sum_{i=1}^{\nu} \log \left(1+ \frac{\lambda_i P_i}{N_{0}}\right),
\label{eq:dataRate}
\end{equation}


\[ 
\text{s.t.}\,\,\,\, tr\left[\boldsymbol{\Sigma}^2    \left(\widetilde{\boldsymbol{V}}\widetilde{\boldsymbol{V}}^{H}\right)^{-1}  \right] = P_{max};
\]

$$
P_i \geqslant 0, \, \, i = 1, \cdots, \nu.
$$

Eq.\ref{eq:dataRate} is a convex problem, the optimal solution satisfying the Karush-Kuhn-Tucker (KKT) conditions is resolvable. We can address this optimization problem via a water-filling algorithm. The optimal solution $P_i$ can be find as,
\begin{equation}
 P_i = max\left(\frac{1}{\mu  \cdot \left[  \left(\widetilde{\boldsymbol{V}}\widetilde{\boldsymbol{V}}^{H}\right)^{-1}\right]_{ii}} - \frac{N_0}{\lambda_i}, \quad 0\right),
\label{eq:result1}
\end{equation}
where $\mu$ is a non-negative Lagrange multiplier deduced from the derivation of the Lagrangian expression:
\begin{equation}
\mu=\frac{\nu}{P_{\max }+\sum_{i=1}^{\nu} \frac{N_{0} \cdot\left[\left(\widetilde{V} \widetilde{V}^{H}\right)^{-1}\right]_{i i}}{\lambda_{i}}}.
\end{equation}

The received power $P_{Q}$ at a random position $Q$ which is in proximity to the BS, is computed as:

\begin{equation}
P_Q =|\boldsymbol{H}^{Q}  \boldsymbol{B}|^{2}.
\end{equation}

In this mechanism, as we only consider the transmit power constraint, there may be several transmit beams which exceed the EMFE threshold out of the safety circle.  In the sequel, we briefly introduce two EMF-aware MU-MIMO BF schemes previously proposed in \cite{ourRIS2022} and then we will detail a novel Dual-GD based EMF-aware MU-MIMO BF scheme that improve the capacity performance while satisfying the EMFE constraints.


\section{EMF-aware MU-MIMO Beamforming in RIS-aided 6G networks}
\label{sec:beamforming}

In this section, we propose a Dual-GD based EMF-aware BF scheme for wireless MU-MIMO DL communications, taking into account the power and EMF constraints. The general problem is described as follows:
\begin{equation}
\mathrm{C}^{*}:=\max _{P_{1}, \ldots, P_{\nu}} \omega \sum_{i=1}^{\nu} \log \left(1+ \frac{\lambda_i P_i}{N_{0}}\right),
\label{eq:dataRate1}
\end{equation}

\begin{equation*}
    \text { s.t. }{tr}\left[\boldsymbol{\Sigma}^{2}\left(\widetilde{\boldsymbol{V}} \widetilde{\boldsymbol{V}}^{H}\right)^{-1}\right] \leqslant P_{max};
\end{equation*}
\begin{equation*}
    P_{Q}= tr\left[\boldsymbol{\Sigma}^2( \boldsymbol{H}^{Q} \widetilde{\boldsymbol{V}}^+ )^H(\boldsymbol{H}^{Q} \widetilde{\boldsymbol{V}}^+)\right] \leqslant \text{EMF}_{th}, \,\, Q \in \Omega;
\end{equation*}
$$
P_i \geqslant 0, \, \, i = 1, \cdots, \nu.
$$
where $\Omega$ is the set of all sampling positions $Q$ on the safety circle.

\subsection{Reduced EMF-aware BF Scheme $\boldsymbol{B}_{red}$}

The reduced EMF-aware BF scheme is carried out by using a reduction in the total transmit power of the reference BF. The corresponding reduction factor $\alpha$ is determined by, 
\begin{equation}
\alpha=min(\frac{\text{EMF}_{\text {th }}  }{\max\limits_{Q \in \Omega}(\mathbf{P}_Q)},1),
\end{equation}
where  
$\mathbf{P}_Q$ is the received power at a sampling position $Q \in \Omega$.

Consequently, for the reduced EMF-aware BF, the transmit power per layer is reduced by this factor $\alpha$, the power allocation matrix is given by $\boldsymbol{\Sigma}_{red} = \sqrt{\alpha} \boldsymbol{\Sigma} $ and the total transmit power is equal to $P_{red}  = tr\left[\boldsymbol{\Sigma}_{red}^2    \left(\widetilde{\boldsymbol{V}}\widetilde{\boldsymbol{V}}^{H}\right)^{-1}  \right] $.

By this way, the reduced EMF-aware BF scheme,  denoted as $\boldsymbol{B}_{red}$, fulfills the EMF exposure constraints at the expense of some network capacity and is given by:
\begin{equation}
    \boldsymbol{B}_{red} = \widetilde{\boldsymbol{V}}^{+}  \boldsymbol{\Sigma}_{red}= \sqrt{min(\frac{\text{EMF}_{\text {th }}  }{\max\limits_{Q \in \Omega}(\mathbf{P}_Q)},1)}\cdot \widetilde{\boldsymbol{V}}^{+}  \boldsymbol{\Sigma}.
\end{equation}

\subsection{Enhanced EMF-aware BF Scheme $\boldsymbol{B}_{enh}$}
\label{sec:truncated}

An enhanced EMF-aware BF scheme $\boldsymbol{B}_{enh}$ was proposed in \cite{ourRIS2022}. The key idea is to evaluate the contribution of each layer to the received power over the safety circle sampling points and to selectively reduce the power of each layer in an iterative way.

We sample $N_Q$ points on the safety circle and calculate the received power $P_Q$ at these $N_Q$ points. First, let $\boldsymbol{B}_{enh} = \boldsymbol{B}$. Then, at each iteration we find the location $Q_{max}$ with the highest received power $P_{Q_{max}}$ and detect the layer $i_0$ which has the greatest influence on its received power. A reduction factor $\beta_{i_0} = (\text{EMF}_{\text{th}}/P_{Q_{max}})$ is applied to reduce the power allocated to layer $i_0$. This iteration is repeated until the received power at all sampled points satisfies the EMFE limits. For more details on this BF scheme, please refer to \cite{ourRIS2022}.


\subsection{Dual Gradient Descent EMF-aware BF Scheme $\boldsymbol{B}_{gd}$}
\label{sec:dual}

A Dual-GD based EMF-aware BF scheme is designed to accommodate the transmit power constraint and the EMFE limits of all $N_Q$ sampling points on the safety circle. Reinforcement learning has been widely applied to optimal decision making for various engineering problems. In particular, the Dual-GD approach addresses the challenge of optimization problems under inequality constraints.

The diagonal power allocation matrix of the proposed Dual-GD BF scheme is denoted by $\boldsymbol{\Sigma}_{gd}$. The Lagrangian function $\mathcal{L}$ is defined as:

\begin{equation}
\mathcal{L}(\boldsymbol{\Sigma}, \boldsymbol{\mu})= \omega \sum\limits_{i=1}^{\nu} \log \left(1+\frac{\lambda_{i} P_{i}}{N_{0}}\right) 
-\sum_{k=0}^{N_Q} \mu_{k} \mathcal{F}_{c,k}(\bm{\Sigma}),
\end{equation}
where $\boldsymbol{\mu} = [\mu_0, \cdots, \mu_{N_Q}]$ is the Lagrangian multipliers for the $(N_Q+ 1)$ conditions and
$\mathcal{F}_{c,k}$ is the function of the $k$-th constraint. When $k= 0$, it refers to the maximum transmit power constraint, so we have,

\begin{equation}
\mathcal{F}_{c,0}(\bm{\Sigma}) =\operatorname{tr}\left[\boldsymbol{\Sigma}^{2}\left(\widetilde{\boldsymbol{V}} \widetilde{\boldsymbol{V}}^{H}\right)^{-1}\right]-P_{\max }. 
\end{equation}
Since we sampled a total number of $N_Q$ points uniformly on the safety circle, $\mathcal{F}_{c,k}$ indicates the EMF limit of the $k^{th}$ observation point with $k \in [1, N_Q]$, 

\begin{equation}
\mathcal{F}_{c,k} (\bm{\Sigma})=\operatorname{tr}\left[\boldsymbol{\Sigma}^{2}( \boldsymbol{H}^{Q_k} \widetilde{\boldsymbol{V}}^+ )^H(\boldsymbol{H}^{Q_k} \widetilde{\boldsymbol{V}}^+)\right]-\text{EMF}_{\text {th }}.
\end{equation}
Then, the Lagrange dual function $g$ is defined as:

\begin{equation}
g(\boldsymbol{\mu})=\mathcal{L}\left({\boldsymbol{\Sigma}}^*, \boldsymbol{\mu}\right) \text{ where } {\boldsymbol{\Sigma}}^*=\arg \max _{\boldsymbol{\Sigma}} \mathcal{L}(\boldsymbol{\Sigma}, \boldsymbol{\mu})
\end{equation}

In this way, we integrate multiple constraints into a single Lagrangian function.
Since the utility function is convex, the strong duality will often hold which means that the minimum value of $g$ equals the maximum value of the optimization problem. Hence, if we find the vector $\bm{\mu}$ that minimizes $g$, we solve the optimization problem.

We initialize the vector $\boldsymbol{\mu}$ to a random value and then we alternate between maximizing the Lagrangian function $\mathcal{L}$ with respect to the primal variables ${\boldsymbol{\Sigma}}$ and then decrement the Lagrange multiplier $\boldsymbol{\mu}$ by its gradient. By repeating the iteration  described by the following three steps, the solution will converge:
\begin{enumerate}
    \item Find $\boldsymbol{\Sigma}^* = \arg \max _{\boldsymbol{\Sigma}} \mathcal{L}(\boldsymbol{\Sigma}, \boldsymbol{\mu})$. The diagonal elements of $\bm{\Sigma}^*$ are computed as follows: 
    \begin{equation}
    \begin{array}{l}
    P_{i}^{*}=\\ \frac{1}{\left[\begin{array}{l}
    \mu_{0}\left(\widetilde{V} \widetilde{V}^{H}\right)_{i i}^{-1}+ \\
    \sum_{k=1}^{N_{Q}} \mu_{k}\left[\left(H^{Q_{k}} \widetilde{V}^{+}\right)^{H}\left(H^{Q_{k}} \widetilde{V}^{+}\right)\right]_{i i}
    \end{array}\right]}-\frac{N_{0}}{\lambda_{i}}.
    \end{array}
    \label{eq:dualmax}
    \end{equation}
    with $1 \leq i \leq \nu$.

    \item Given the value of ${\boldsymbol{\Sigma}}^*$, the gradient descent step with respect to each Lagrangian multiplier  ${\mu_k}$ is calculated: 
    \begin{equation}
    \Delta_k = \frac{d g\left(\boldsymbol{\mu}\right)}{d \mu_{k}}=\frac{d \mathcal{L}\left({\boldsymbol{\Sigma}}^{*}, \boldsymbol{\mu}\right)}{d \mu_{k}}.
    \label{eq:dual2}
    \end{equation}
    Note that for the first transmit power constraint ($k=0$), $\Delta_0$ is given by:  
    $$\Delta_0 = -\sum_{i=1}^{\nu} P_{i}\left(\widetilde{V}  \widetilde{V}^{H}\right)_{i i}^{-1} + P_{max};$$

For the $N_Q$ other constraints corresponding to the EMFE constraints at the different safety circle locations, the expression of  $\Delta_k$ is given by :
	$$\Delta_k = -\sum_{i=1}^{\nu} P_{i} \left[( \boldsymbol{H}^{Q_k} \widetilde{\boldsymbol{V}}^+ )^H(\boldsymbol{H}^{Q_k} \widetilde{\boldsymbol{V}}^+)\right]_{ii}+\text{EMF}_{\text {th }}.$$

    \item Update each Lagrangian multiplier  by its gradient, 
    \begin{equation}
    \mu_{k}=  \mu_{k}-\beta_k \cdot \Delta_{k},
    \label{eq:dual3}
    \end{equation}
    where $\beta_k$ is the learning rate for the Lagarange multiplier $\mu_k$.

\end{enumerate}

With the updated values of $\boldsymbol{\mu} = [\mu_0, \cdots, \mu_i]$, repeat the gradient descent process mentioned in steps (1) - (3). When all the condition functions, $k\in[0,\cdots, N_Q]$, satisfy $\mathcal{F}_{c,k} \leqslant \tau$, with $\tau \geqslant 0$ being a predefined tolerance threshold, it means  that the solution converges. The final $\bm{\Sigma}^*$ corresponds to the Dual-GD power allocation denoted by  solution ${\boldsymbol{\Sigma}_{gd}}$. Consequently, the Dual-GD EMF-aware BF scheme is written as:

\begin{equation}
\begin{aligned}
\boldsymbol{B}_{gd} =\widetilde{\boldsymbol{V}}^{+}  \boldsymbol{\Sigma}_{gd}.
\end{aligned}
\end{equation}
The detailed Dual-GD algorithm is disciplined in Algorithm \ref{alg:2}.

\begin{algorithm}
	\renewcommand{\algorithmicrequire}{\textbf{Input:}}
	\renewcommand{\algorithmicensure}{\textbf{Output:}}
	\caption{Dual-GD EMF-aware BF}
	\label{alg:2}
	\begin{algorithmic}[1]
		\STATE 
     	 Sample $N_Q$ points uniformly on the safety circle as observation points; \\

     	\STATE Initialize all the Lagrange multipliers $\boldsymbol{\mu}$ with random positive values and  $\boldsymbol{\Sigma}_{gd}$ denotes the BF matrix of the Dual-GD BF scheme.
     	
     	\STATE Set the learning rates $\beta_k \in (0,1)$ $\forall{ k\, = 0, \cdots, N_Q}$ for all $N_Q+1$ constraints;
     	
     	\STATE Set the  tolerance threshold $\tau = 10^{-3}$;

		\WHILE{$\exists \, k\in [0, \cdots, N_Q]$ s.t. $\mathcal{F}_{c,k}(\bm{\Sigma_{gd}}) \geqslant \tau$}

		\STATE Calculate $\boldsymbol{\Sigma_{gd}} = \arg \max _{\boldsymbol{\Sigma}} \mathcal{L}(\boldsymbol{\Sigma}, \boldsymbol{\mu})$ as shown in eq.\ref{eq:dualmax} with the given values of $\boldsymbol{\mu}$;\\

		\STATE Compute the derivative  $\Delta_k= \frac{d g\left(\boldsymbol{\mu}\right)}{d \mu_{k}}$ of each $\mu_k$ with  $\boldsymbol{\Sigma_{gd}}$ obtained in the previous step.

		\FOR{$0 \leqslant k \leqslant N_Q$}
		
		\IF{$\mathcal{F}_{c,k}\geqslant \tau$}
		
		\STATE $\mu_{k}=\mu_{k}-\beta_k \cdot \Delta_{k}$

		\ENDIF
		
		\ENDFOR

		\ENDWHILE
	    \RETURN $\boldsymbol{\Sigma}_{gd}$ and $\boldsymbol{B}_{gd}$ 
	\end{algorithmic} 
	\label{alg:truncated}
\end{algorithm}

\section{Numerical results}
\label{sec:numerical}
In this section, we numerically evaluate the performance of the Dual-GD EMF-aware BF scheme and compare it to the reduced and enhanced EMF-aware BF schemes introduced in \cite{ourRIS2022}. Assume that the BS is equipped with a 2D antenna array with $8 \times 8$ pairs of cross-polarized antenna elements (total of 128 antennas elements).  The height of the BS is $25$ m. There are $Z=3$ RIS and $S = 3$ scatterers randomly distributed in the cellular network. Each RIS has $K = 4$ antenna elements.  We assume that  $L = [3,4,5,6,7,8,9]$ numbers of UEs with random positions are allocated in the cell and each UE has $N=4$ antennas. The heights of the RIS, the scatterers and the UEs are all equal to $1.5$ m.  
The maximum transmit power of the BS is $P_{max} = 200$ Watt. We set the radius of the safety circle to $R = 50$ m. The EMF-threshold is $\text{EMF}_{th} = 52$ dBm. We limit $\nu_l=2$ spatial streams per user for transmission. 
The carrier frequency is assumed to be $3.5$ GHz and with a channel bandwidth of $100$ MHz.

 \begin{figure*}[htbp]
 \centering
   \includegraphics[width=0.9\textwidth,height=0.48\textwidth]{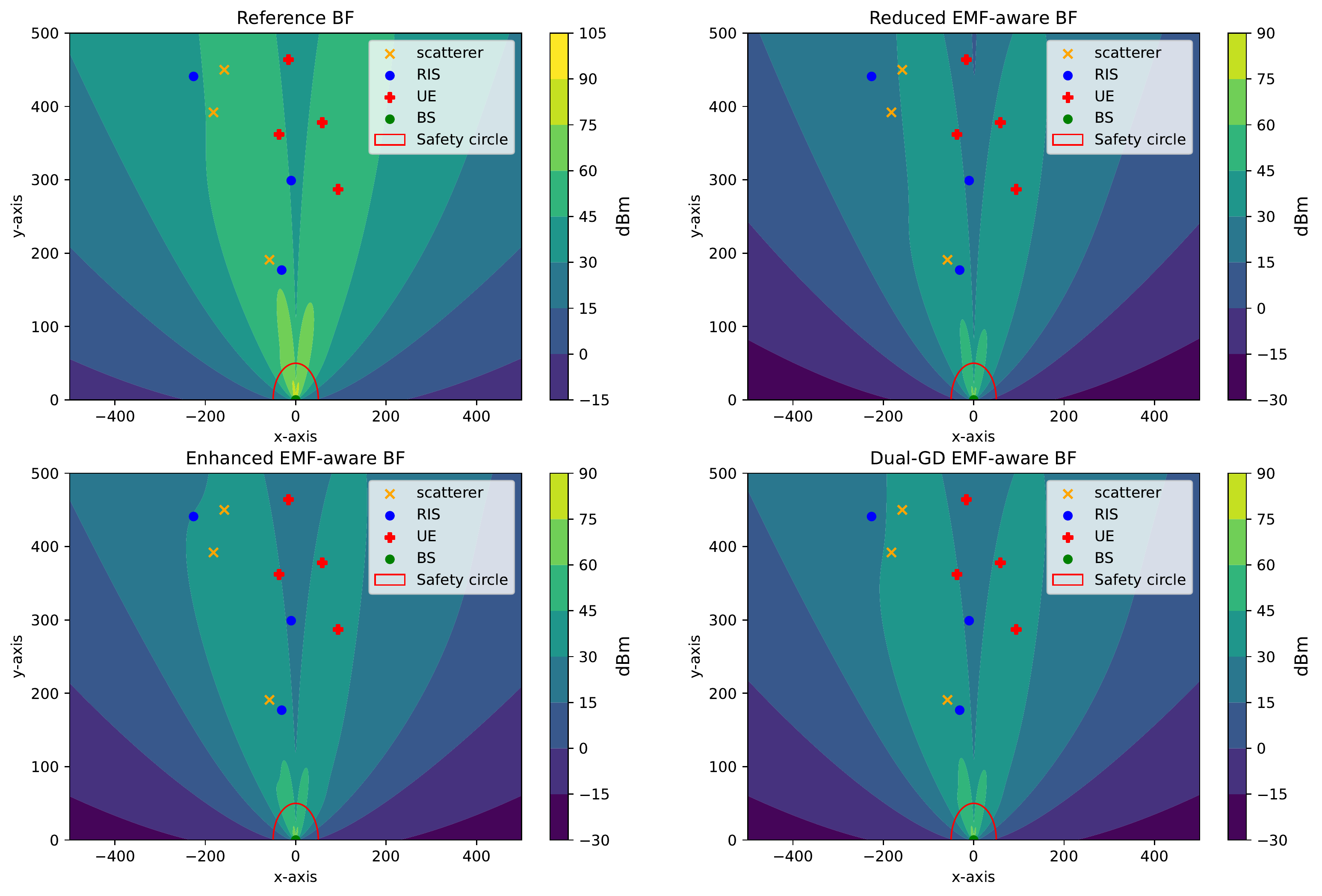} 
           \caption{The received power distribution in a given space with different BF schemes}
               \label{fig:Pr_four}
 \end{figure*} 

Figure \ref{fig:Pr_four} shows the received power in a given observation space for $L=4$ UEs. In reference BF case, the received power in the given space ranges from $-15$ dBm to $105$ dBm. By using the reduced, the enhanced and the Dual-GD EMF-aware BF schemes respectively, the distribution of the received power over the whole given area has been significantly changed. The maximum received power in the given space has been reduced to $90$ dBm for those three EMF-aware BF schemes.

 \begin{figure*}[t]
 \centering
   \includegraphics[width=0.8\textwidth,height=0.45\textwidth]{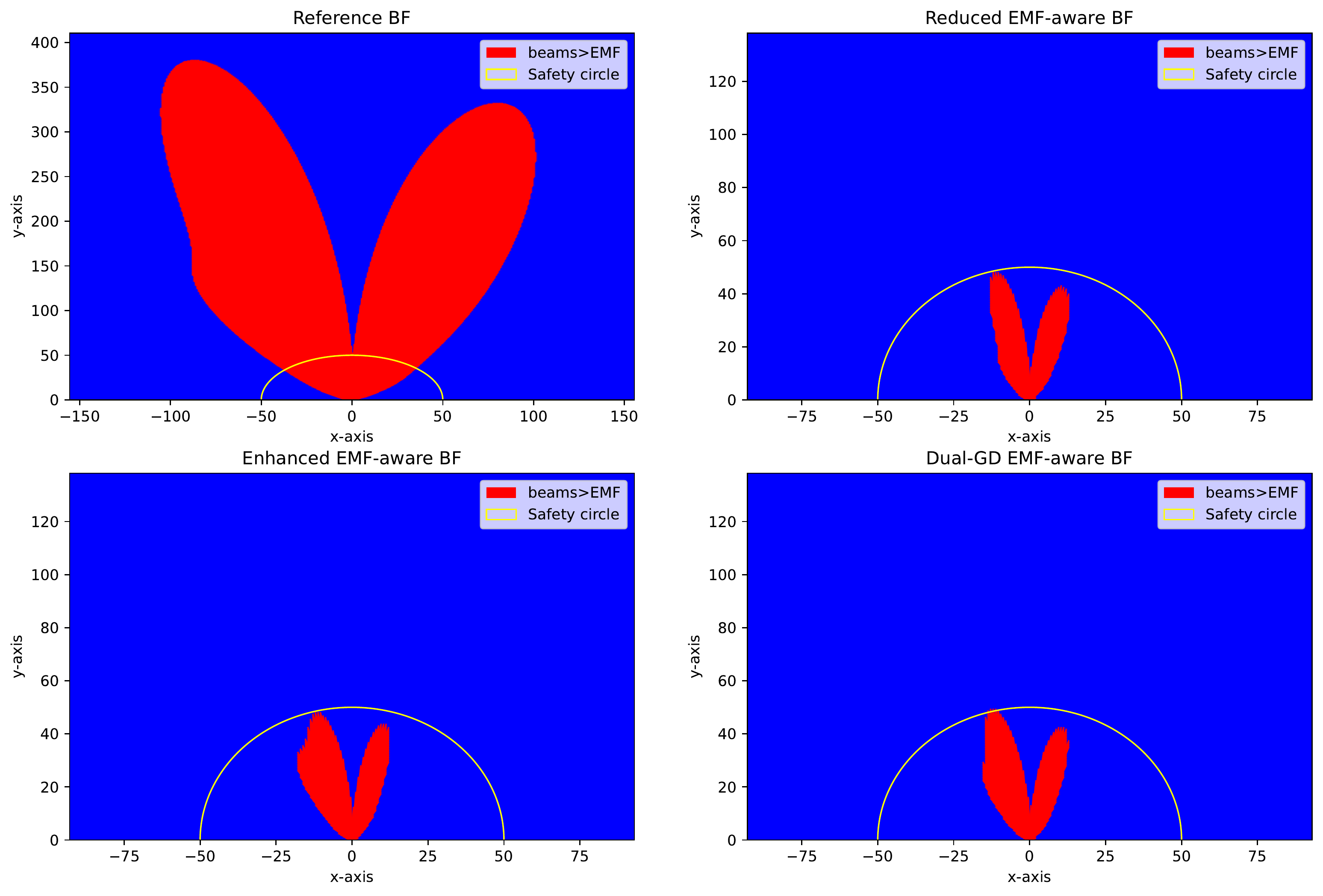} 
           \caption{Beams exceed the EMFE constraints in the given space}
               \label{fig:Pr_four_binary}
 \end{figure*} 

Figure \ref{fig:Pr_four_binary} is the illustration of beams that exceed the EMFE threshold in the same scenario as shown in figure \ref{fig:Pr_four}. 
In the reference case, there are multiple beams that exceed the EMFE limits beyond the safety circle. In all three EMF-aware BF cases, the EMFE constraint is well adhered in the open space outside the safety circle. The reduced EMF-aware achieves this goal by decreasing the overall power by a given factor. In the enhanced algorithm, the transmit power of the different layers is modified in such a way that the exact EMFE limits are achieved at the safety circle points that correspond to the exceeding directions. By adopting the Dual-GD EMF-aware BF, we adjust the transmit power per layer through the gradient descent of the Lagrange multipliers. In this way, we take into account the impact of each constraint and decrease or increase the transmit power per layer intelligently.  In Fig. \ref{fig:Pr_four_binary}, the difference of the shape of exceeding beams also confirms the difference between the three EMF-aware BF schemes.

\begin{table}[]
\centering
\begin{tabular}{ |c||c|c|c|  }
\hline
\multicolumn{4}{|c|}{SINR per layer (dB) } \\
\hline
Layers& Reduced  & Enhanced & Dual-GD\\
\hline
1 & $65.02$ & $63.74$ & $64.57$\\
2 & $67.73$ & $68.20$ & $68.99$\\
3 & $61.72$ & $66.11$ & $67.12$\\
4 & $66.64$ & $69.55$ & $68.62$\\
5 & $35.38$ & $35.89$ & $36.18$\\
6 & $40.18$ & $40.80$ & $41.46$\\
7 & $58.89$ & $63.40$ & $63.01$\\
8 & $65.32$ & $63.54$ & $62.23$\\
\hline
\end{tabular}
\vspace{0.4cm}
\caption{SINR per layer for three different EMF-aware BF schemes}
\label{tab:sinr}
\end{table}

Table \ref{tab:sinr} provides the SINR values for each layer of the three EMF-aware BF schemes under the same scenario as Fig. \ref{fig:Pr_four} and Fig. \ref{fig:Pr_four_binary}. 
Since these three algorithms adjust the transmit power of each layer in different ways, the final received SINR of each layer also varies with the transmit power. The highest total SINR on the receiver sides is given by the Dual-GD EMF-aware BF scheme with $74.30$ dB. One can observe that the Dual-GD scheme can achieve an important per-layer SINR gain compared to the "reduced" BF scheme: up to $5$ dB of SINR gain. 

\begin{figure}[t]
\centering
\includegraphics[width=0.41\textwidth,height=0.35\textwidth]{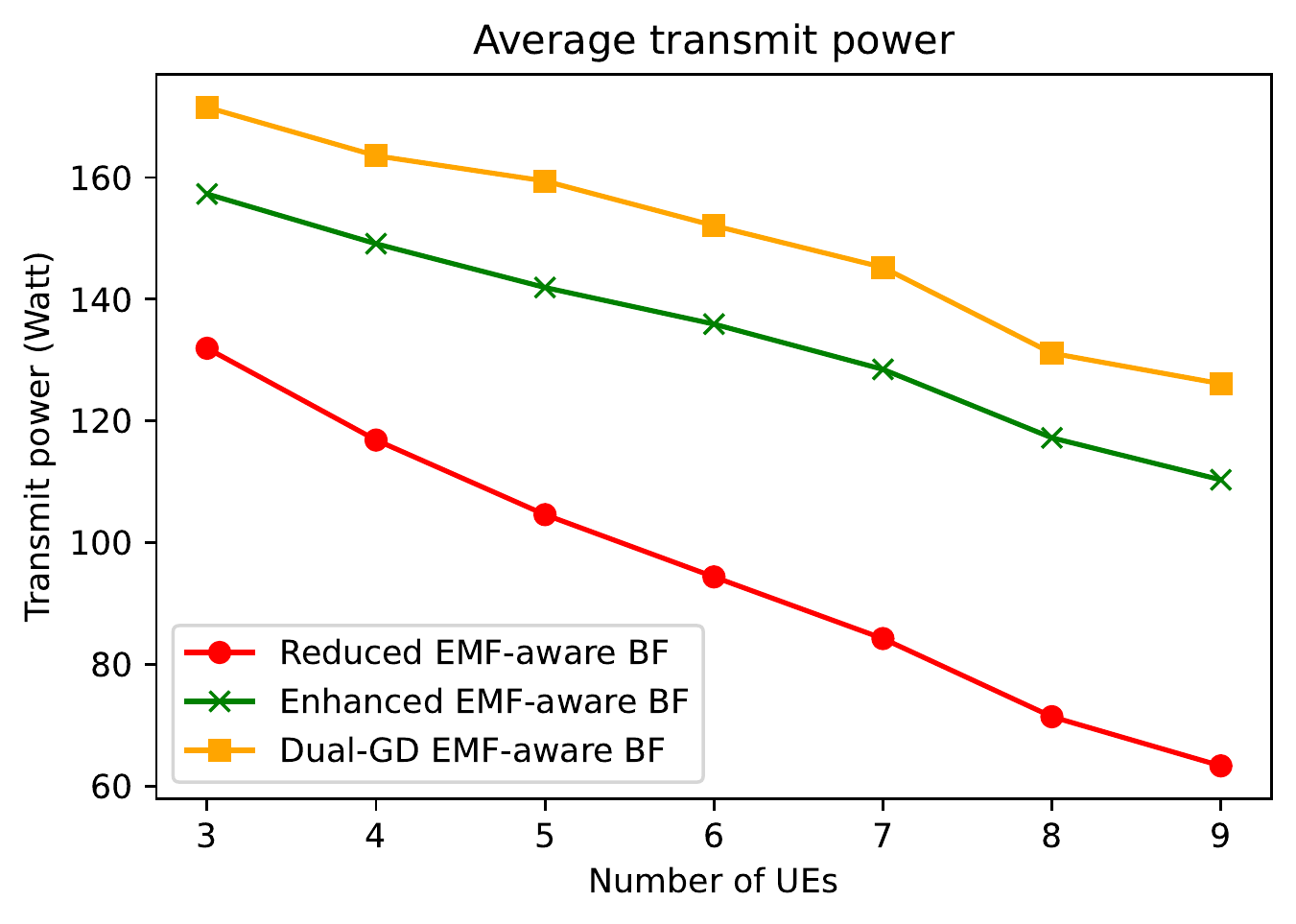} 
\caption{Average transmit power of different EMF-aware BF schemes}
\label{fig:average_Pi}
\end{figure}   

\begin{figure}[htbp]
\centering
\includegraphics[width=0.4\textwidth,height=0.35\textwidth]{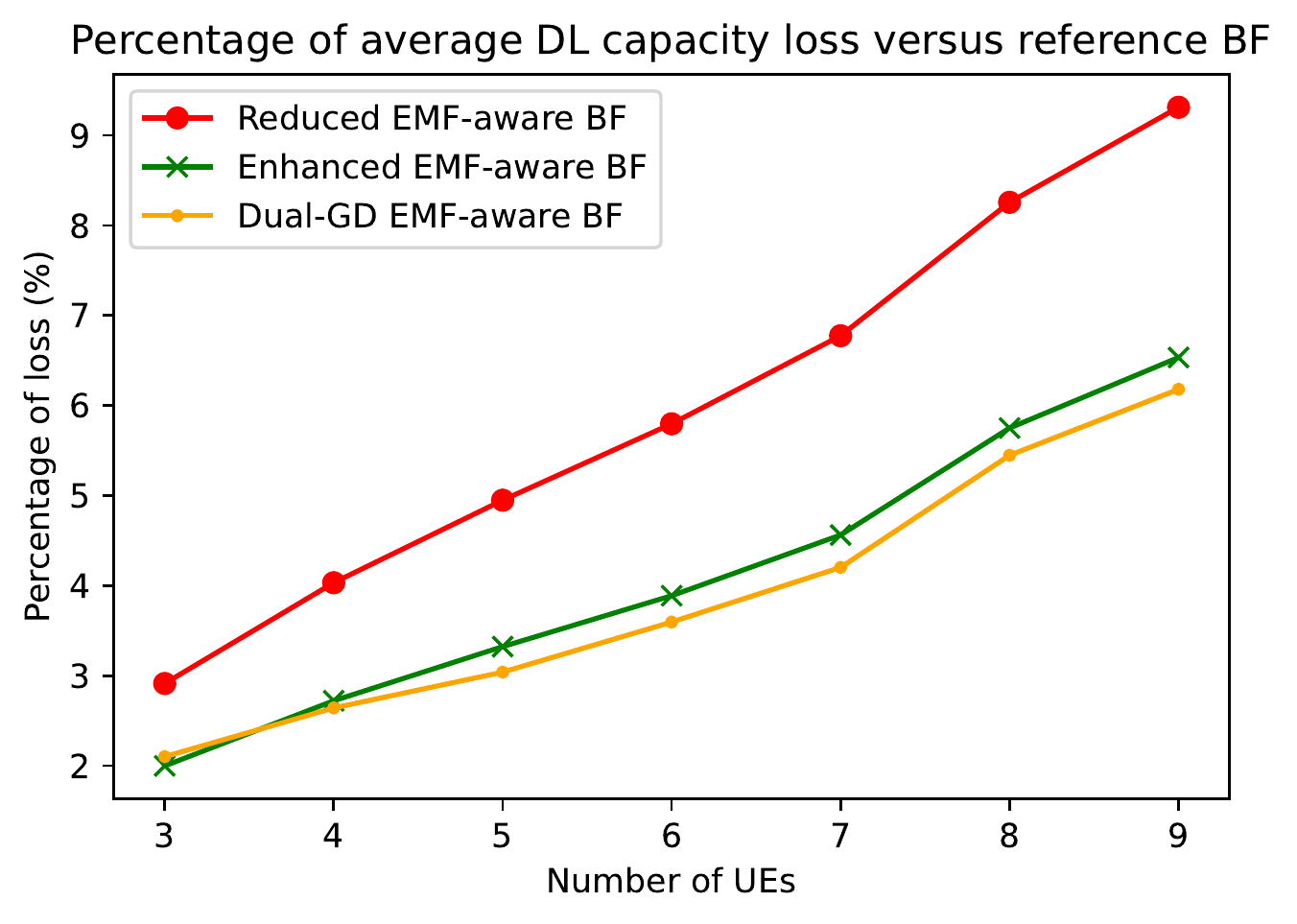} 
\caption{The average DL capacity loss of the cellular network compared to reference BF case}
\label{fig:average_c}
\end{figure}


Moreover, we evaluate the performance of the proposed BF scheme considering various number of UEs, i.e. from $L=3$ to $9$. We also consider $200$ samples of channels with different locations of UEs, scatterers and RISs corresponding to each different number of $L$.

Figure \ref{fig:average_Pi} presents the average transmit power at the BS for the three EMF-aware BF schemes. As shown in the figure, the Dual-GD EMF-aware BF can still guarantee the EMF constraints at a higher transmit power compared to the other two BF modes. Its tolerated transmit power is about 8\% higher than the enhanced BF, and up to about 120\% higher than the reduced BF scheme. That is, when operators need to transmit data with high power for practical reasons, the Dual-GD BF scheme provides the maximum possibility to ensure that the EMFE in the observation area is not exceeded.
Figure \ref{fig:average_c} plots the percentage of average capacity loss of the DL communication relative to the reference case. 
As this figure demonstrates, Dual-GD EMF-aware BF achieves the lowest capacity loss. With strict control of transmit power, Dual-GD loses no more than $6\%$ of the network capacity.

\section{Conclusion}
\label{sec:conclusion}

In this paper, we modeled the  DL communcation for RIS-aided MU-MIMO systems considering the latest 3GPP antenna pattern.
A novel Dual-GD BF scheme is proposed to address EMFE regulation. We also compare the simulation performance of this new BF scheme with the two other EMF-aware BF schemes proposed previously. The Dual-GD EMF-aware BF scheme is able to meet EMF constraints at higher transmit power with less loss of system capacity than the other two BF schemes.
In the near future, we will jointly optimize the transmit precoding weight and the power allocation scheme in order to achieve higher performance while satisfying EMFE limits.

\section{Acknowledgement}
\label{sec:ack}

This work was conducted within the framework of the European Union's innovation project RISE6G.

\begin{appendices}

\section{3GPP antenna pattern}
\label{app:antenna}

According to recent 3GPP release \cite{3GPPRelease17},
the vertical cut of the radiation pattern is,

\begin{equation}
A_{\mathrm{dB}}\left(\theta, \phi=0^{\circ}\right)= -\min \left\{12\left(\frac{\theta-90^{\circ}}{\theta_{3 \mathrm{dB}}}\right)^{2}, S L A_{V}\right\},
\end{equation}
where $\theta_{3 \mathrm{dB}}=65^{\circ}$, $S L A_{V}=30 \mathrm{~dB}$ and $\theta \in\left[0^{\circ}, 180^{\circ}\right]$.
So the 3D (total)  radiation pattern for an antenna element is:

\begin{equation}
\begin{array}{l}
A_{\mathrm{dB}}\left(\theta, \phi\right)=  \\ 8 -\min \left\{-\left(A_{\mathrm{dB}}\left(\theta, \phi=0^{\circ}\right)+  
A_{\mathrm{dB}}\left(\theta=90^{\circ}, \phi\right)\right), A_{\max }\right\}.
\end{array}
\end{equation}

From the individual antenna element's antenna pattern, we can derive the 3D radiation pattern $A_{\text {Beam }}(\theta, \varphi)$ of the entire antenna array as described below:
\begin{equation}
A_{\text {Beam }}(\theta, \varphi)=A_{dB}(\theta, \varphi)+10 \log _{10}\left(\left|\sum_{m=1}^{M}  w_{m} \right|^{2}\right),
\end{equation}
where $w_{m}=\frac{1}{\sqrt{N_H}} \exp \left(-j \frac{2 \pi}{\lambda}(m\%N_H-1) d_{V} \cos \theta_{\text {etilt }}\right)$ is the complex weight with a pre-tilt  angle $\theta_{\text {etilt }}$.

In case of polarized antennas, the polarization is modeled as angle-independent in both azimuth and elevation. In the horizontal polarization, the antenna element field component is given by:
\begin{equation}
F_{\theta,\phi}=\sqrt{A_{\text {Beam }}(\theta, \varphi)} \sin (\zeta),
\end{equation}
with $\zeta = +/- 45^{\circ}$ being the polarization slant angle corresponds to a pair of cross-polarized antenna elements.

\section{Calculation of channel gains}
\label{app:derivation}

As mentioned in section \ref{sec:model}, $G_{m, U_{n}^{l}}$, $G_{m, s, U_{n}^{l}}$ and $ G_{m, R_{k}^{z}, U_{n}^{l}} $ are the channel gains with respect to different propagation paths $m  \rightarrow U_{n}^{l}$, $m \rightarrow s \rightarrow U_{n}^{l}$ and $m \rightarrow R^z_k \rightarrow U_{n}^{l}$. Since the scatterers, users, RIS and BS are assumed to be far away from each other, one can apply the planar wave approximation to the corresponding channels. They are calculated as:
\begin{equation}
G_{m, U_{n}^{l}}=F^{\prime}_{\theta,\phi}\cdot \sigma \cdot e^{-j \frac{2 \pi}{\lambda}\varsigma(m, U_{n}^{l})} ;
\end{equation}
\begin{equation}
G_{m, s, U_{n}^{l}}=F^{\prime}_{\theta,\phi}\cdot \beta(s) \cdot e^{-j \frac{2 \pi}{\lambda}\left(\delta(m, s)+\delta\left(s, U_{n}^{l}\right)\right)} ;
\end{equation}
\begin{equation}
\begin{aligned}
& G_{m, R_{k}^{z}, U_{n}^{l}}  =\\& F^{\prime}_{\theta,\phi} \cdot  r^{\text {ris }} \cdot \epsilon\left(R_{0}^{z}\right) \cdot e^{-j \frac{2 \pi}{\lambda} \cdot \eta\left(m, R_{k}^{z}\right)} \cdot \mathbf{w}_{k}^{z} \cdot e^{-j \frac{2 \pi}{\lambda} \eta\left(R_{k}^{z}, U_{n}^{l}\right)};
\end{aligned}
\end{equation}
where $F^{\prime}_{\theta,\phi}$ is the 3GPP radiation power pattern corresponds to the spherical angles $(\theta,\phi)$ of a given path and $F^{\prime}_{\theta,\phi}$ is converted to a linear scale.
$\sigma$, $\beta\left(s\right)$ and $\epsilon(R^{z}_0)$ are complex random Gaussian variables with unit expectation. $r^{r i s}=1 / K$ is the refection amplitude and  $\mathbf{w}^{z} \in \mathbb{C}^{K \times 1}$ is the RIS BF reflection weight. 
In addition, 

\begin{equation}
\varsigma(m, U_{n}^{l})=\frac{\overrightarrow{A_{m}^{BS}  A_{l, 0}^{UE}}}{\left\|\overrightarrow{A_{m}^{BS}  A_{l, 0}^{UE}}\right\|}\cdot \overrightarrow{A_{l, 0}^{UE}  A_{l, n}^{U E}};
\end{equation}

\begin{equation}
\delta(m, s)=\frac{\overrightarrow{A_{0}^{BS} A_{s}^{s c a}}}{\left\|\overrightarrow{A_{0}^{BS}  A_{s}^{s c a}}\right\|} \cdot \overrightarrow{A_{0}^{BS}  A_{m}^{B S}};
\end{equation}

\begin{equation}
\delta\left(s, U_{n}^{l}\right)=\frac{\overrightarrow{A_{s}^{s c a}  A_{l, 0}^{UE}}}{\left\|\overrightarrow{A_{s}^{s c a}  A_{l, 0}^{UE}}\right\|} \cdot \overrightarrow{A_{l, 0}^{UE}  A_{l, n}^{U E}};
\end{equation}
\begin{equation}
\eta\left(m, R_{k}^{z}\right)=\frac{\overrightarrow{A_{0}^{B S} A_{z, 0}^{RIS}}}{\left\|\overrightarrow{A_{0}^{BS}  A_{z, 0}^{RIS}}\right\|} \cdot\left(\overrightarrow{A_{0}^{BS}  A_{m}^{BS}}+\overrightarrow{A_{z, 0}^{RIS} A_{z, k}^{RIS}}\right);
\end{equation}
\begin{equation}
\eta\left(R_{k}^{z}, U_{n}^{l}\right)=\frac{\overrightarrow{A_{z, 0}^{RIS} A_{l, 0}^{U E}}}{\left\|\overrightarrow{A_{z, 0}^{RIS}  A_{l, 0}^{U E}}\right\|}\cdot \left(\overrightarrow{A_{z, 0}^{RIS}  A_{z, k}^{RIS}}+\overrightarrow{A_{l, 0}^{U E}  A_{l, n}^{U E}}\right).
\end{equation}
with $A^{BS}_0$ being the center position of the BS linear array, 
$A^{BS}_m\in \mathbb{R}^{3 \times 1}$ is the position of the $m^{th}$ BS antenna element. Similarly, $A^{UE}_{l,n}\in \mathbb{R}^{3 \times 1}$ is the position of the $n^{th}$ antenna element of the $l^{th}$ UE and $A^{RIS}_{z,k}\in \mathbb{R}^{3 \times 1}$ is the position of the $k^{th}$ element of the $z^{th}$ RIS. The position of scatterer $s$ is denoted as $A^{sca}_{s}\in \mathbb{R}^{3 \times 1}$.
\end{appendices}

\bibliographystyle{IEEEtran}
\bibliography{ref.bib}

\end{document}

%% file: macro.tex
\definecolor{red}{rgb}{0.9,0.1,0.1}
\definecolor{purple}{rgb}{0.625,0.125,0.9375}
\definecolor{brown}{rgb}{0.5,0.3,0.1}
\definecolor{orange}{rgb}{1,0.4,0}
\definecolor{yellow}{rgb}{0.9,0.8,0.2}
\definecolor{green}{rgb}{0.1,0.75,0.1}
\definecolor{cyan}{rgb}{0.1,0.8,0.8}
\definecolor{lblue}{rgb}{0.1,0.4,0.8}
\definecolor{blue}{rgb}{0.1,0.1,0.9}
\definecolor{magenta}{rgb}{0.9,0,0.8}
\definecolor{mgreen}{rgb}{0.2,0.5,0.2}
\definecolor{gray}{rgb}{0.5,0.5,0.5}
\definecolor{reliefcolor}{gray}{0.8}
\definecolor{emphcolor}{rgb}{1,.5,1}
\definecolor{emphcolorr}{rgb}{1,.9,1}
\definecolor{emphcolorb}{rgb}{.9,.9,1}
\definecolor{emphcolorg}{rgb}{.9,1,.9}
